\documentclass{article} 
\usepackage{color}
\usepackage{amsmath,amsfonts,amsthm,amssymb}
\usepackage{graphicx}
\usepackage{fullpage}

\numberwithin{equation}{section}

\newtheorem{theorem}{Theorem}

\RequirePackage{dsfont}

\renewcommand{\vec}[1]{\mathbf{#1}}

\newcommand{\remove}[1]{}

\newcommand{\F}{\mathbb{F}}
\newcommand{\E}{\mathbb{E}}

\newcommand{\abs}[1]{\left| #1 \right|}

\newcommand{\poly}{{\rm poly}}

\renewcommand{\P}{{\rm P}}
\newcommand{\NP}{{\rm NP}}
\newcommand{\RP}{{\rm RP}}
\newcommand{\BP}{{\rm BP}}
\newcommand{\UP}{{\rm UP}}
\newcommand{\PH}{{\rm PH}}
\newcommand{\parP}{{\ensuremath{\oplus {\rm P}}}}
\newcommand{\numP}{{\ensuremath{\# {\rm P}}}}

\title{A simple constant-probability \RP\ reduction from \NP\ to \parP}

\author{Cristopher Moore \\
 \textsf{moore@cs.unm.edu}\\
 Department of Computer Science\\
 University of New Mexico \\
 and the Santa Fe Institute
 \and
 Alexander Russell\\
 \textsf{acr@cse.uconn.edu}\\
 Department of Computer Science and Engineering\\
 University of Connecticut
}

\begin{document}
\maketitle 

\begin{abstract}
The proof of Toda's celebrated theorem that the polynomial hierarchy is contained in $\P^\numP$ relies on the fact that, under mild technical conditions on the complexity class $\mathcal{C}$, we have $\exists \,\mathcal{C} \subset \BP \cdot \oplus \,\mathcal{C}$.  More concretely, there is a randomized reduction which transforms nonempty sets and the empty set, respectively, into sets of odd or even size. The customary method is to invoke Valiant's and Vazirani's randomized reduction from \NP\ to \UP, followed by amplification of the resulting success probability from $1/\poly(n)$ to a constant by combining the parities of $\poly(n)$ trials. Here we give a direct algebraic reduction which achieves constant success probability without the need for amplification.  Our reduction is very simple, and its analysis relies on well-known properties of the Legendre symbol in finite fields.
\end{abstract}

Valiant and Vazirani~\cite{valiant-vazirani} gave a clever randomized reduction from \NP\ to \UP, the class of promise problems which have either a unique solution or no solution at all.  Their reduction works as follows.  Given, say, a 3-SAT formula $\phi$ on $n$ variables, we begin choose an integer $k$ uniformly from $\{1,\ldots,n\}$.  We then add the additional constraint that a hash function $h$ takes the value zero, where $h$ is chosen from a pairwise independent family of hash functions, and where a given truth assignment $\vec{x}$ obeys $h(\vec{x})=0$ with probability $2^{-k}$.  If $\phi$ is satisfiable, then with probability $\Omega(1/n)$ this additional constraint makes the solution unique.

So long as the complexity class $\mathcal{C}$ is expressive enough to compute the hash function $h$ and is closed under intersection, this reduction asserts that:
\[
\exists\, \mathcal{C} \subseteq \RP_{\poly} \cdot \exists! \, \mathcal{C}\,,
\]
where $\RP_{\poly}$ denotes one-sided error with a $1/\poly(n)$ probability of success and where $\exists!$ denotes unique existence.  Since $1$ is odd, we can also write
\[
\exists\, \mathcal{C} \subseteq \RP_{\poly} \cdot \oplus \,\mathcal{C}\,,
\]
where, for instance, $\parP$ is the class of decision problems which ask whether the number of witnesses for a problem in \NP\ is odd.

For the case of $\parP$, we can amplify the probability of success as follows: if we perform $m=\Omega(n)$ independent trials of this reduction, then with probability $\Omega(1)$ at least one trial will yield a formula $\phi'$ with a unique solution (assuming $\phi$ is satisfiable).  Since the expression
\[
a = 1 + \prod_{i=1}^m (a_i+1)
\]
is odd if and only if at least one of the $a_i$ is odd, and it is easy to implement such expressions within $\parP$ by constructing $m$-tuples of witnesses, we conclude
\[
\exists \,\textrm{P}\subseteq \RP \cdot \oplus\,\textrm{P}
\]
where now the reduction works with probability $\Omega(1)$.  (Of course, by taking, say, $m=n^2$, we can make the probability of success exponentially close to $1$.)
By showing that the operators $\BP$ and $\oplus$ can be commuted, we obtain Toda's result~\cite{toda} that 
\[
\PH \subseteq \BP \cdot \oplus \P \subseteq \P^\numP \, . 
\]

The purpose of this note is to give an alternate reduction from \NP\ to $\RP^{\parP}$
which works with constant probability without the need for amplification.  Our reduction is quite simple, and may be of independent interest.  First, let $p$ be a prime, let $\F_p$ denote the field of order $p$, and for $a \in \F_p$ let $\chi(a)$ denote the Legendre symbol
\[
\chi(a) = \begin{cases} 
0 & \mbox{if $a=0$} \\
+1 & \mbox{if $a=b^2$ for some $b \ne 0$} \\
-1 & \mbox{otherwise} \, . 
\end{cases}
\]
If $p$ has $\poly(n)$ digits, then $\chi(a)$ can be computed in polynomial time as follows.  Using modular exponentiation, calculate
\[
t = a^{(p-1)/2} \bmod p \, .
\]
Then $\chi(a) = +1$ or $-1$ if $t = 1$ or $p-1 \equiv -1$ respectively.

Now consider the following theorem.
\begin{theorem}
\label{thm:odd}
Let $S$ be a nonempty subset of $\F_p$ of size $\abs{S} = o(p^{1/2})$.  If $b$ is chosen uniformly at random from $\F_p$, then the set
\[
S' = \{ x \in S \mid \chi(x+b) = -1 \}
\]
is of odd size with probability $1/2-o(1)$.
\end{theorem}

\begin{proof}
First note that, with probability $1-|S|/p = 1-o(n^{-1/2})$, we have $x+b \ne 0$ for all $x \in S$.  Henceforth we will assume that this is the case.

Then note that $S'$ is of odd size if and only if 
\[
\prod_{x \in S} \chi(x+b) = -1 \, .
\]
Since $\chi$ is a \emph{multiplicative character}, i.e., since $\chi(ab) = \chi(a) \chi(b)$, we can write this as
\[
\chi\!\left( \prod_{x \in S} (x+b) \right) = -1 \, .
\]
Then
\[
\Pr\left[ \mbox{$\abs{S'}$ is odd} \right] = \frac{1-T}{2}
\; \mbox{ where } \;
T = \E_b \,\chi\!\left( \prod_{x \in S} (x+b) \right) \, .
\]

Now note that $\prod_{x \in S} (x+b)$ is a polynomial function of $b$.  The expectation of a multiplicative character on the image of a polynomial on $\F_q$ is bounded by the following theorem, proved by A. Weil:

\begin{theorem}[\cite{weil}]
Let $\chi$ be a multiplicative character of $\F_p$ of order $m > 1$ (that is, $m$ is the least integer for which $\chi(a)^m=1$ for any $a$).  Let $f(b) \in \F_p[x]$ be a polynomial that is not the $m$th power of a polynomial, and let $d$ be the number of distinct roots of $f$ in its splitting field over $\F_q$.  Then
\[
\abs{ \sum_{b \in \F_p} \chi(f(b)) } \le (d-1) \,p^{1/2} \, .
\]
\end{theorem}

In our case, $m=2$ and $f(b) = \prod_{x \in S} (x+b)$.  Since this product gives a complete factorization of $f(b)$ into distinct linear terms, $f(b)$ is certainly not the square of a polynomial.  Moreover, it has degree $|S|$, so $d \le |S| = o(p^{1/2})$. Therefore, 
\[
T = \frac{1}{p} \sum_{b \in \F_p} \chi(f(b)) 
= o(1)
\]
and $\abs{S'}$ is odd with probability $(1-T)/2 = 1/2-o(1)$.
\end{proof}
See also~\cite[\S5]{lidln} for further discussion.

Our reduction works as follows.  For concreteness, suppose we have a 3-SAT formula $\phi$ on $n$ variables.  Choose a prime $p > 2^{cn}$ for some $c > 2$, so that $2^n = o(p^{1/2})$.  Interpret each truth assignment $\vec{x}$ as an $n$-bit integer $x$, choose $b$ uniformly from $\F_p$, and add the constraint that $\chi(x+b) = -1$.  Then by Theorem~\ref{thm:odd}, if $\phi$ is satisfiable, the resulting formula $\phi'$ will have an odd number of satisfying assignments with probability $1/2-o(1)$.


\end{document}